\documentclass[pra,twocolumn,superscriptaddress,amsmath,amssymb, showpacs]{revtex4}

\usepackage{graphicx}
\usepackage{dcolumn}
\usepackage{bm}

\begin{document}



\title{Dynamical control of two-level system's decay and long time freezing}

\author{Wenxian Zhang}


\affiliation{The State Key Laboratory for Advanced Materials and Devices,
Department of Optical Science and Engineering, Fudan University, Shanghai
200433, China}

\affiliation{Ames Laboratory and Iowa State University, Ames, Iowa 50011, USA}

\author{Jun Zhuang}

\affiliation{The State Key Laboratory for Advanced Materials and Devices,
Department of Optical Science and Engineering, Fudan University, Shanghai
200433, China}

\date{\today}

\begin{abstract}
We investigate with exact numerical calculation coherent control of a two-level
quantum system's decay by subjecting the two-level system to many periodic
ideal $2\pi$ phase modulation pulses. For three spectrum intensities (Gaussian,
Lorentzian, and exponential), we find both suppression and acceleration of the
decay of the two-level system, depending on difference between the spectrum
peak position and the eigen frequency of the two-level system. Most
interestingly, the decay of the two-level system freezes after many control
pulses if the pulse delay is short. The decay freezing value is half of the
decay in the first pulse delay.
\end{abstract}


\maketitle

\section{Introduction}

It is of fundamental interests to study the quantum control of
spontaneous emission which inspired the development of the quantum
electrodynamics during the last century~\cite{Milonni94, Scully97,
Allen75}. The unwanted spontaneous emission often sets the
ultimate limit of precise quantum measurement and many proposals
have been made to suppress it~\cite{Frishman03, Evers02,
Frishman01, Agarwal01}. One way widely applied to control the
spontaneous emission of a small quantum system, such as a neutral
atom, is to place it either into or nearby a microcavity such that
only a single mode or several modes of the cavity are resonant to
the eigenfrequency of the quantum system~\cite{Vogel06}. In this
way the structure of the vacuum is modified and the spontaneous
emission could be controlled by the properties of the cavity.
Another way to control the spontaneous emission is dynamical
control of the quantum system such that the coupling between the
quantum system and the vacuum is effectively modified. For
example, dynamical decoupling of the quantum system from the
vacuum via either phase modulation or amplitude modulation could
in principle extend the coherent time of the system~\cite{Itano90,
Itano91}. Analogous to the quantum Zeno effect (ZE) which says
frequent measurements of a quantum system would prevent the decay
of an unstable quantum system~\cite{Misra77,Itano90, Itano91}, the
extension of the coherent time through coherent modulations of the
quantum system are often called ZE as well. Under some unfavorable
conditions, it is also possible that frequent modulations lead to
acceleration of the spontaneous emission, the so-called quantum
anti-Zeno effect (AZE)~\cite{Kofman00, Agarwal01a, Fischer01,
Kofman01}.

In practice people employ both ways separately or combination of
them to realize the optimal control of a quantum system. G. S.
Agarwal {\it et al.}'s work~\cite{Agarwal01, Agarwal01a} is among
many of such works. In their paper, a two-level model system with
both structured vacuum and free space vacuum are investigated.
They found significant suppression of spontaneous emission rate
for structured vacua but either suppression or acceleration may
appear in a free space vacuum, depending on the frequency of the
control pulses which is $2\pi$ phase modulation pulses. They
demonstrated for a free space vacuum that ZE shows up for
$\omega_0\tau=1$ while AZE appears for $\omega_0\tau=\pi$ where
$\omega_0$ is the eigenfrequency of the two-level system and
$\tau$ is the delay time between control pulses. At the end they
argued that AZE is possible for
$\tau>\omega_0^{-1}$~\cite{Agarwal01}.

A puzzle arises if one accepts the AZE condition because neither ZE or AZE
appear if one leaves the quantum system alone in which case obviously $\tau \gg
\omega_0^{-1}$. In this paper we revisit this problem by adopting the exact
solution of a two-level quantum system which does not require the weak coupling
and short time approximations compared to Ref.~\cite{Agarwal01}. By
investigating several typical structured vacuum, we show the conditions for
quantum ZE and AZE and the boundary between them.

Another puzzle is the ZE effect of a large number of pulses.
According to the ZE, at a fixed evolution time $t$, the survival
probability of the initially excited state approaches 1 with
infinite number of pulses $N\rightarrow \infty$ (the pulse delay
$\tau=t/N$ approaches 0). It is unclear what the road map looks
like as $N$ increases. Two ways might be possible. One way is that
the rate of decay depends solely on the pulse delay $\tau$
(independent on $N$) and approaches zero if $\tau\rightarrow
0$~\cite{Scully03}. Another way is that the decay rate depends
only on the number of pulses $N$ (independent on $\tau$) and the
decay rate becomes zero after initial several pulses. For the
latter one, we expect to see decay freezing at large number of
pulses. We will find which way is the correct one in this paper.

The paper is organized as follows. Section~\ref{sec:rabi} reviews
briefly Rabi oscillation of a two-level system under a
perturbation and Sec.~\ref{sec:crabi} develops an exact formula of
Rabi oscillation under periodic pulses. In Sec.~\ref{sec:qza}
after establishing a quasi-level picture of the pulsed two-level
system with constant spectrum intensity, we investigate the
quantum Zeno and anti-Zeno effect for three spectrum intensities,
including the Gaussian, Lorentzian, and exponential one. The
boundary between quantum Zeno and anti-Zeno effect is given. We
discuss the decay freezing as pulse delay $\tau$ getting small in
Sec.~\ref{sec:df}. Finally, conclusion and discussion are given in
Sec.~\ref{sec:cd}.

\section{Rabi oscillations of a two-level system}
\label{sec:rabi}

Let us consider a simple model of a two-level quantum system with
constant coupling to its environment modes. Such a model serves as
the base of further discussions for more complicated systems with
mode-dependent coupling. The Hamiltonian of the system with
detuning $\delta=\omega-\omega_0$ (we set $\hbar=1$ for
convenience) and coupling strength $\nu$ is described
by~\cite{Scully97, dipolenote}
\begin{eqnarray}
H   &=& {\delta\over 2} (|e\rangle\langle e| -
        |g\rangle\langle g|) + \nu (|e\rangle\langle g| +
        |g\rangle\langle e|),
\end{eqnarray}
where $\omega_0=E_e-E_g$ is the difference between eigenfrequencies of the
excited state $|e\rangle$ and the ground state $|g\rangle$, with respectively
eigenenergy $E_e$ and $E_g$, and $\omega$ is the frequency of the external mode
coupled to the two-level system~\cite{picturenote}. Utilizing Pauli matrices, the Hamiltonian can
be rewritten as
\begin{eqnarray}
H &=& {\delta' }  \sigma_z + \nu \sigma_x \label{eq:HPauli}
\end{eqnarray}
with $\sigma_z = |e\rangle\langle e| - |g\rangle\langle g|$,
$\sigma_x = |e\rangle\langle g| + |g\rangle\langle e|$, and
$\delta'=\delta/2$.

The evolution operator of the coupled two-level system is
\begin{eqnarray}
U &=&  d_1 |e\rangle\langle e| + d_1^*|g\rangle\langle g|
    + d_2 (|e\rangle\langle g| + |g\rangle\langle e|)
\end{eqnarray}
where $d_1 = \cos\Omega t - i(\delta'/\Omega) \sin\Omega t$ and
$d_2 = -i (\nu /\Omega)\sin\Omega t$ with
$\Omega^2=\delta'^2+\nu^2$. The transition probability at time $t$
from $|e\rangle$ to $|g\rangle$ is
\begin{eqnarray}
p_{eg}(t) &=& {\nu^2\over \Omega^2} \sin^2(\Omega t)
\end{eqnarray}
providing the initial state is $\psi(0)=|e\rangle$. The system revives at times
such that $\Omega t = k \pi$ with $k$ being an integer.

\section{Controlled Rabi oscillations of a two-level system}
\label{sec:crabi}

\begin{figure}
\includegraphics[width=2.5in]{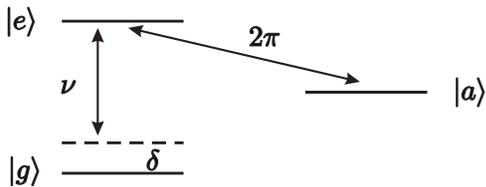}
\caption{Diagram of a coupled two-level system subjected to ideal
$2\pi$ pulses.} \label{fig:levels}
\end{figure}

By applying an ideal $2\pi$ pulse (or parity kick, see
Fig.~\ref{fig:levels}), which is very strong in amplitude and
short in time but gives a $\pi$ phase shift solely to the excited
state $|e\rangle$ utilizing an auxiliary state
$|a\rangle$~\cite{Agarwal01}, the state of the system changes
according to
\begin{eqnarray}
x |e\rangle + y |g\rangle &\underrightarrow{\;\;2\pi\; {\rm pulse}\;\;}& -x
|e\rangle + y |g\rangle
\end{eqnarray}
where $|x|^2+|y|^2=1$. We denote such a pulse as $Z$ pulse hereafter
\begin{eqnarray}
Z(x|e\rangle+y|g\rangle) =-x|e\rangle+y|g\rangle.
\end{eqnarray}
In fact, $Z=-\sigma_z=|g\rangle\langle g| - |e\rangle\langle e|$.

The evolution operator for the $2\pi$ pulse and the free evolution
$\tau$ is
\begin{eqnarray}
U^{(1)} &=& ZU = \left(\begin{array}{cc} d_1 & d_2\\-d_2&-d_1^*
     \end{array} \right).
\end{eqnarray}
For $N$ such operations, the evolution operator at time $t =
N\tau$ becomes
\begin{eqnarray}
U^{(N)} &=& \left(\begin{array}{cc} d_1 &
    d_2\\-d_2&-d_1^*\end{array} \right)^N.
\end{eqnarray}
Let $d_1=d_{1r}+id_{1i}$ and $d_2=id_{2i}$. After some
straightforward simplifications with the use of Pauli matrices,
one easily obtains
\begin{eqnarray}
U^{(N)} &=& i^N (\cos N\lambda - i \sigma_\alpha \sin N\lambda)
\end{eqnarray}
where $\sin\lambda = \sqrt{1-d_{1i}^2}$ and $\sigma_\alpha =
(1/\sin\lambda) (d_{1r} \sigma_z - d_{2i} \sigma_y)$. Note that
$\lambda$ depends on $\tau$ instead of $t=N\tau$.

The transition probability from the initial state $|e\rangle$ to
the ground state $|g\rangle$ is
\begin{eqnarray}
p_{eg}'(t=N\tau) &=& |\langle g| U^{(N)} |e\rangle| ^2 =
p_{eg}(\tau) { \sin^2 N\lambda \over \sin^2\lambda}.
\end{eqnarray}
The above result is exact for any coupling strength and pulse
delay $\tau$. For weak coupling $\nu\ll\delta$, $\Omega\approx
\delta'$ and $\lambda\approx \delta'\tau + \pi/2 +2k\pi$ with $k$
an integer, then
\begin{eqnarray}
p_{eg}' &\approx & {\nu^2 \over \delta'^2}
    \tan^2\delta' \tau \sin^2 N\delta'\tau
\end{eqnarray}
for even $N$, which is exactly Eq. (9) in Agarwal {\it et al.}'s
paper~\cite{Agarwal01}.

\section{Quantum Zeno and anti-Zeno effects}
\label{sec:qza}

The above results are applicable only to single mode bath/environment which
couples to the central two-level system. In general the bath has multimodes and
the coupling may depend on the mode, e.g., the dipolar coupling between an atom
and an electric and magnetic field. For a many-mode bath, the transition
probability of the two-level system subjected to control pulses is in general
given by
\begin{eqnarray}
p_{eg}' &=& \sum_n {\nu^2 \over \Omega_n^2} \sin^2\Omega_n \tau
    {\sin^2 N\lambda_n \over \sin^2\lambda_n}
\end{eqnarray}
with $n$ the bath mode index~\cite{mmnote}. Note that $\Omega_n$ and $\lambda_n$ become mode dependent in the many-mode bath case and $\nu$ is constant here but will be considered as mode dependent in Sec.~\ref{subsec:fdc}. Assuming the bath
spectrum is dense, we turn the summation over mode index $n$ into
the integration over mode frequency $\omega$,
\begin{eqnarray}
p_{eg}' &=& \int d\omega \rho(\omega) {\nu^2 \over \Omega^2}
    \sin^2\Omega \tau
    {\sin^2 N\lambda \over \sin^2\lambda}
\label{eq:cse}
\end{eqnarray}
where $\rho(\omega)$ is the density of states of the bath. For $N=1$ we get the
free decay results (no control pulse). Before considering the real physical
systems which usually have mode dependent coupling, we study several toy models
with constant coupling strength $\nu=\nu_0$ for all bath modes to gain some
ideas about the effect of control pulses.

\subsection{Uniform spectrum intensity}

\begin{figure}
\includegraphics[width=3.25in]{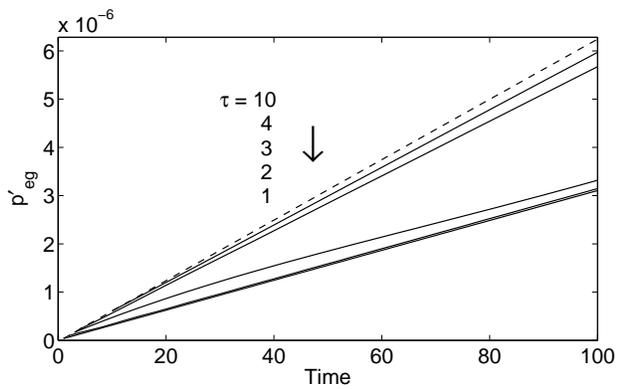}
\caption{Control pulse effect with fixed pulse delays. The parameters are
$\omega_0=1, \nu_0=0.001, \omega_c=100$, and $\tau = 10,4,3,2,1$ from top to
bottom for solid lines. The dashed line denotes the free decay. Time is in units of $1/\omega_0$ hereafter.}
\label{fig:cse1atau}   
\end{figure}

Taking $\rho(\omega) = \rho_0 \equiv 1/\omega_c$ if
$\omega\in[0,\omega_c]$ and $\rho(\omega)=0$ otherwise with
$\omega_c$ denoting the cutoff frequency of the bath,
Fig.~\ref{fig:cse1atau} shows typical free decay and controlled
decay of the excited state. Except at very short times,
Fig.~\ref{fig:cse1atau} shows that the transition probability
linearly depends on the total evolution time. Utilizing the linear
dependence, one defines decay rate (Einstein constant) at long
time $t$ as
\begin{eqnarray}
A &\equiv & {\partial p_{eg}(t) \over \partial t} = 2\pi \rho_0 \nu_0^2
\end{eqnarray}
for free decay~\cite{Agarwal01} and
\begin{eqnarray}
A' &\equiv & {1\over \tau} {\partial p_{eg}' \over \partial N}
    = {2\pi \rho_0\nu_0^2 \over \tau^2} \sum_k
    {1\over \nu_0^2 + (\omega_k-\omega_0)^2/4}
\end{eqnarray}
for controlled decay, where $\omega_k$ is the $k$th resonant mode
frequency~\cite{Kofman01}. For weak coupling $\nu_0 \ll \pi/\tau$, the $k$th
resonant mode lies approximately at $\omega_k \approx \omega_0 +
(2k+1)\pi/\tau$ where $\sin\lambda \ll 1$ and $\sin 2N\lambda /\sin\lambda
\approx \pi\delta(\lambda) \approx (2\pi/\tau)\delta(\omega-\omega_k)$ for
large $N$. Note that the two nearest neighbor peaks around $\omega_0$ with
$k=0$ and $k=-1$ contribute equally about $40\%$ among all the
peaks~\cite{Agarwal01, Kofman01}.

\begin{figure}
\includegraphics[width=3in]{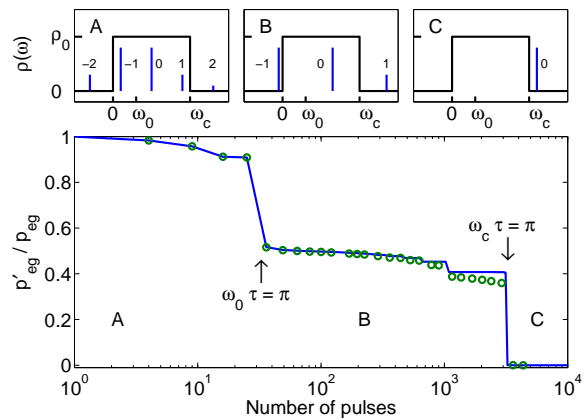}
\caption{(Color online) Control pulse effect {\it vs} number of pulses $N$
(main panel) and relative positions of resonant modes to $\omega_0$ and
$\omega_c$ (upper three panels A, B, and C corresponds to three regions in the
main panel, respectively). The parameters are $N\tau = 100, \omega_0=1,
\nu_0=0.001$, and $\omega_c=100$. Two big steps shows two major peaks out of
the lower and upper cutoff frequency, respectively. These small steps shows
minor peaks going out of the cutoff frequencies. Circles are for the 
frequency-dependent coupling case.}
\label{fig:cse1a}   
\end{figure}

If we assume the bath has only positive frequency and concentrate
on the two dominant peaks, we find
\begin{eqnarray}
{A'\over A} &\simeq & \left\{\begin{array}{cc} 0,& \omega_c \tau<\pi \\
    {1/2},& \omega_0 \tau<\pi<\omega_c \tau \\
    1,&{\rm otherwise}\end{array}\right.
\end{eqnarray}
and $p_{eg}'/p_{eg}$ also shows step-like behavior as depicted in
Fig.~\ref{fig:cse1a}. Moreover, Fig. \ref{fig:cse1a} exhibits only quantum Zeno
effect, i.e., suppression of the decay by the control pulses. The decay are
completely suppressed, $A'=0$, once the control pulse frequency is larger than
the cutoff frequency.

\subsection{Gaussian spectrum intensity}

A Gaussian spectrum intensity has a form as
\begin{eqnarray}
\rho(\omega) &=& {1\over \sqrt{2\pi} \;\Gamma}
    \;e^{-(\omega-\omega_m)^2/2\Gamma^2}, \; \; \omega\ge 0
\end{eqnarray}
where $\Gamma$ and $\omega_m$ denotes the width and the position of the maximal
intensity, respectively.

\subsubsection{$\omega_0=\omega_m$}

\begin{figure}
\includegraphics[width=3.25in]{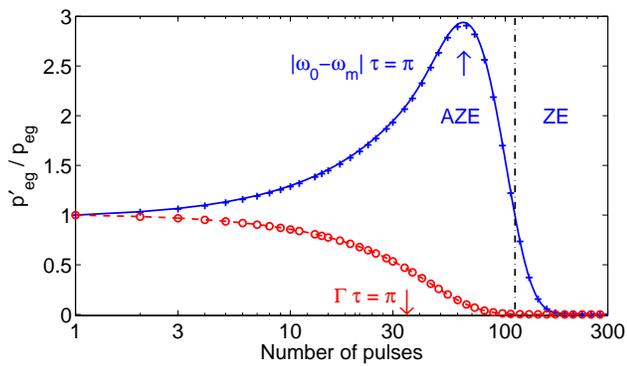}
\caption{(Color online) Reduced transition probability at fixed time ($t =
N\tau =100$) {\it vs} number of pulses $N$ for Gaussian spectrum intensity for
$\omega_0=\omega_m$ (red dashed line) and $\omega_0=\omega_m-2$ (blue solid
line). Circles and crosses are for the corresponding frequency-dependent
coupling case. The vertical dot-dashed line shows the boundary between quantum
Zeno and anti-Zeno effects for $\omega_0=\omega_m-2$. Other parameters are
$\nu_0=0.001, \omega_c=100, \omega_m=\omega_c/2$, $\Gamma = 1$.}
\label{fig:cse1bii}   
\end{figure}

In this case we expect only quantum Zeno effect to appear, $A'\le A$,
because it is easy to check that
\begin{eqnarray}
{A'\over A} &=& {4\over \pi^2} \sum_k {1\over (2k+1)^2} \;
    {\rho(\omega_k) \over \rho(\omega_0)} \le 1
\end{eqnarray}
where we have used the large $N$ and weak coupling assumptions. By considering
the two dominant peaks, we further obtain that
\begin{eqnarray}
{A' \over A} &\approx & {8\over \pi^2} \;
    e^{-\pi^2 / 2 \Gamma^2 \tau^2}
\label{eq:g0}
\end{eqnarray}
for small $\tau$ such that $\Gamma\tau \lesssim \pi$. Equation~(\ref{eq:g0})
shows that the decay rate decreases rapidly with the control pulse frequency in
a Gaussian form. For exceedingly small $\tau$ which satisfies $\Gamma\tau \ll
\pi$, $A'$ is essentially zero, which means the transition is inhibited and the
survival probability of the initial state saturates. The red dashed line in
Fig.~\ref{fig:cse1bii} demonstrates the quantum Zeno effect and the prohibition
of the transition.

\subsubsection{$|\omega_0-\omega_m|\gtrsim \Gamma$}

For $|\omega_0-\omega_m| < \Gamma$, the results are similar to
$\omega_0 = \omega_m$ and exhibit only quantum Zeno effect. To
observe substantially enhancement of the transition, i.e., the
quantum anti-Zeno effect \cite{Kofman00},
$|\omega_0-\omega_m|\gtrsim \Gamma$ is required. More
specifically, by taking the biggest peak around $\omega_0$, one
obtains the necessary condition for quantum anti-Zeno effect as
\begin{eqnarray}
|\omega_0-\omega_m| \ge \sqrt{4\ln {\pi\over 2}} \;\Gamma \approx
1.3 \Gamma.
\end{eqnarray}

The blue solid line in Fig.~\ref{fig:cse1bii} shows the quantum Zeno and
anti-Zeno effects for $|\omega_0-\omega_m|=2$. Both quantum Zeno (large $N$
region) and anti-Zeno (small $N$ region) effects are observed. At fixed time
$t=N\tau$, large $\tau$ (small N) gives anti-Zeno effect while small $\tau$
(large N) gives Zeno effect. The boundary between quantum Zeno and anti-Zeno
effect is about $2|\omega_m -\omega_0|\tau \approx \pi$ if $|\omega_m
-\omega_0|\gg \Gamma$. Strongest anti-Zeno effect is obtained at
$|\omega_0-\omega_m| \tau = \pi$ where one of the two main resonant modes lies
near $\omega_m$.

\subsection{Lorentzian and exponential spectrum intensity}

\begin{figure}
\includegraphics[width=3.25in]{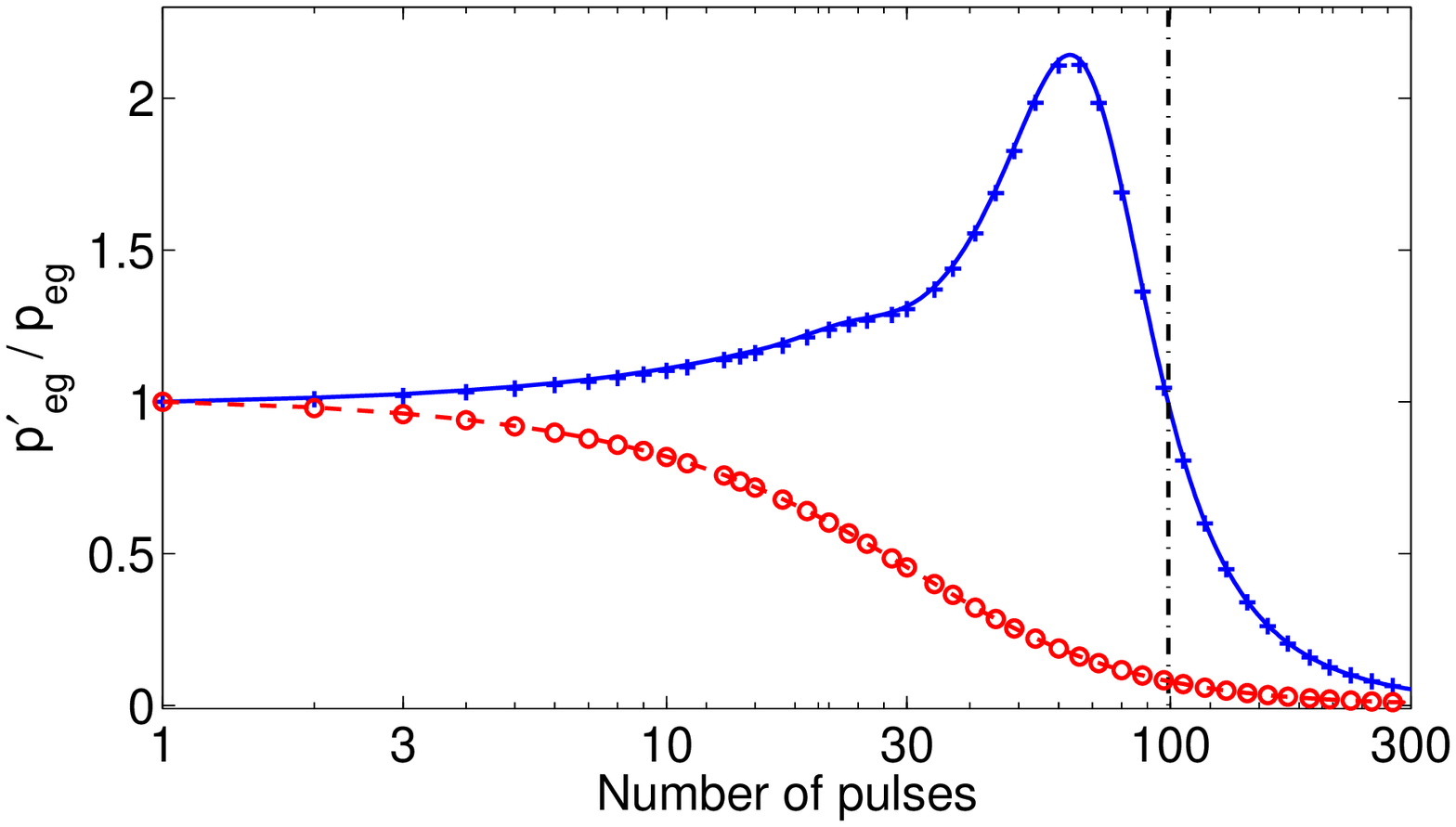}
\includegraphics[width=3.25in]{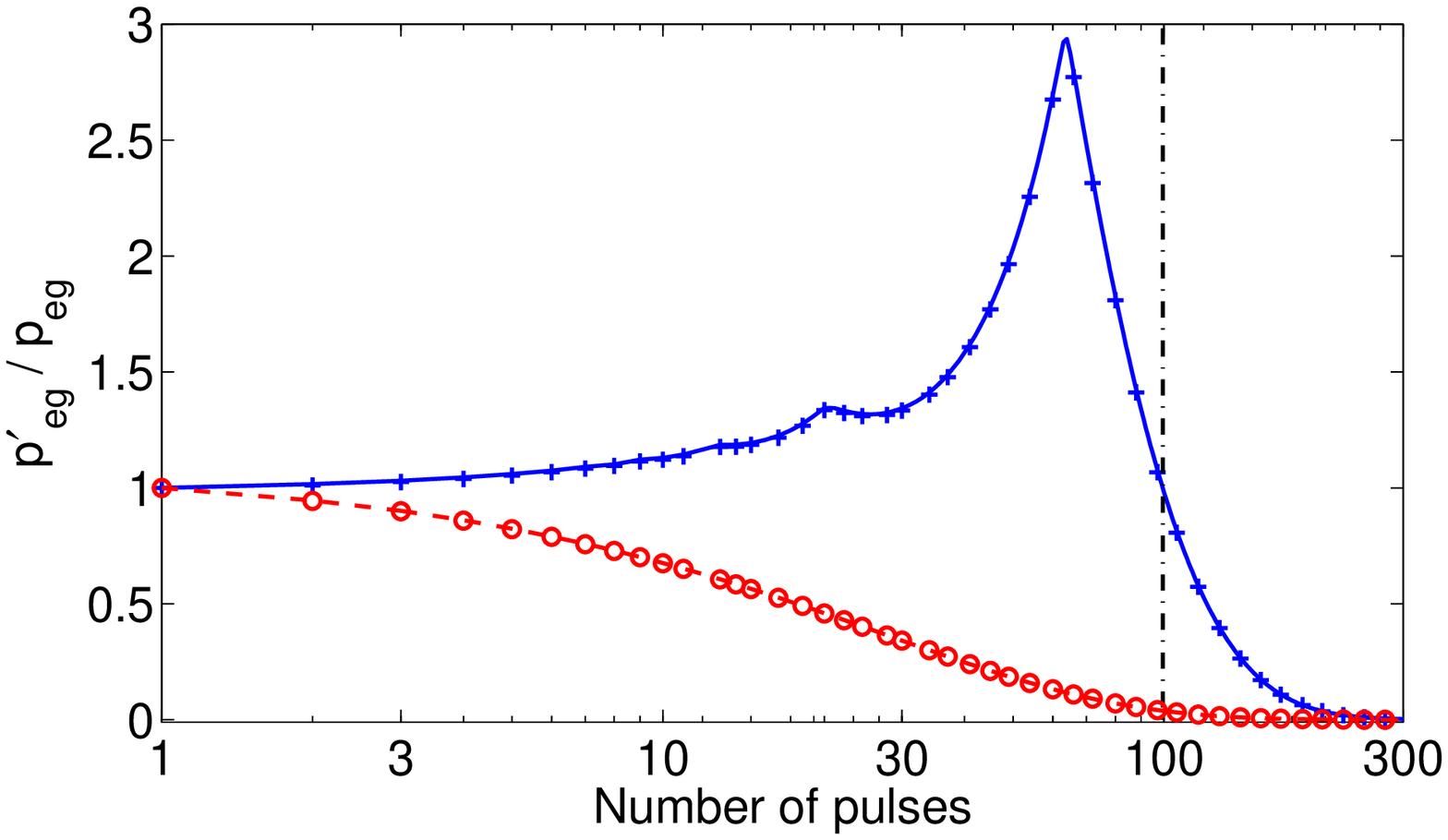}
\caption{(Color online) Same as Fig. \ref{fig:cse1bii} except that the spectrum
intensity is Lorentzian (top) and exponential (bottom).}
\label{fig:cse1cdii}   
\end{figure}

The spectrum intensity for Lorentzian and exponential are,
respectively,
\begin{eqnarray}
\rho(\omega) &=& {\Gamma/\pi \over (\omega-\omega_m)^2 +
\Gamma^2}, \; \omega \ge 0  \\
\rho(\omega) &=& {1\over 2\Gamma} e^{-|\omega-\omega_m|/\Gamma}, \; \omega \ge
0  .
\end{eqnarray}
Similar to Gaussian spectral intensity, only the quantum Zeno effect is
observed if $\omega_0=\omega_m$ for the Lorentzian and exponential shape
(Fig.~\ref{fig:cse1cdii}). The transition probability is essentially inhibited
once $\tau \ll \tau_c$ where $\tau_c\sim 1/\omega_c$. Once
$|\omega_0-\omega_m|\gtrsim \Gamma$, we find both quantum Zeno and anti-Zeno
effects and the boundary between Zeno and anti-Zeno regime is determined
approximately by $2|\omega_m-\omega_0|\tau\approx \pi$. The peak position of
the quantum anti-Zeno effect lies about at $|\omega_0-\omega_m|\tau \approx
\pi$. As shown in Fig.~\ref{fig:cse1cdii}, the reduced transition probability
in the exponential case has a narrower peak width than that in the Lorentzian
case. Two small secondary peaks are noticeable in the anti-Zeno region of the
exponential spectrum case. These peaks are due to the second and third
quasi-level resonant to $\omega_m$. They satisfy the condition, respectively,
$|\omega_0-\omega_m|\tau \approx 3\pi, 5\pi$.

\subsection{Frequency dependent coupling}
\label{subsec:fdc}

The widely adopted dipolar coupling in spontaneous emission of a
two level atom or molecule has a frequency dependence as
$\nu(\omega) = \nu_0 \sqrt{\omega}$ where $\nu_0$ is taken as a
constant, which depends on the dipole matrix elements
\cite{Scully97}. We will consider the same coupling spectrum
intensity as before, i.e.,
\begin{eqnarray}
\rho(\omega) &=& {1\over \omega} {1\over \sqrt{2\pi} \;\Gamma}
    \;e^{-(\omega-\omega_m)^2/2\Gamma^2}, \; \omega \ge 0 \\
\rho(\omega) &=& {1\over \omega} {\Gamma/\pi \over
(\omega-\omega_m)^2 +
\Gamma^2}, \; \omega \ge 0 \\
\rho(\omega) &=& {1\over \omega} {1\over 2\Gamma}
e^{-|\omega-\omega_m|/\Gamma}, \; \omega \ge 0
\end{eqnarray}
for Gaussian, Lorentzian, and exponential density of state, respectively. As
shown in Figs.~\ref{fig:cse1bii} and~\ref{fig:cse1cdii},
the frequency dependent coupling has little effect on the
performance of the control pulses.

\section{Decay freezing}
\label{sec:df}

\begin{figure}
\includegraphics[width=3.25in]{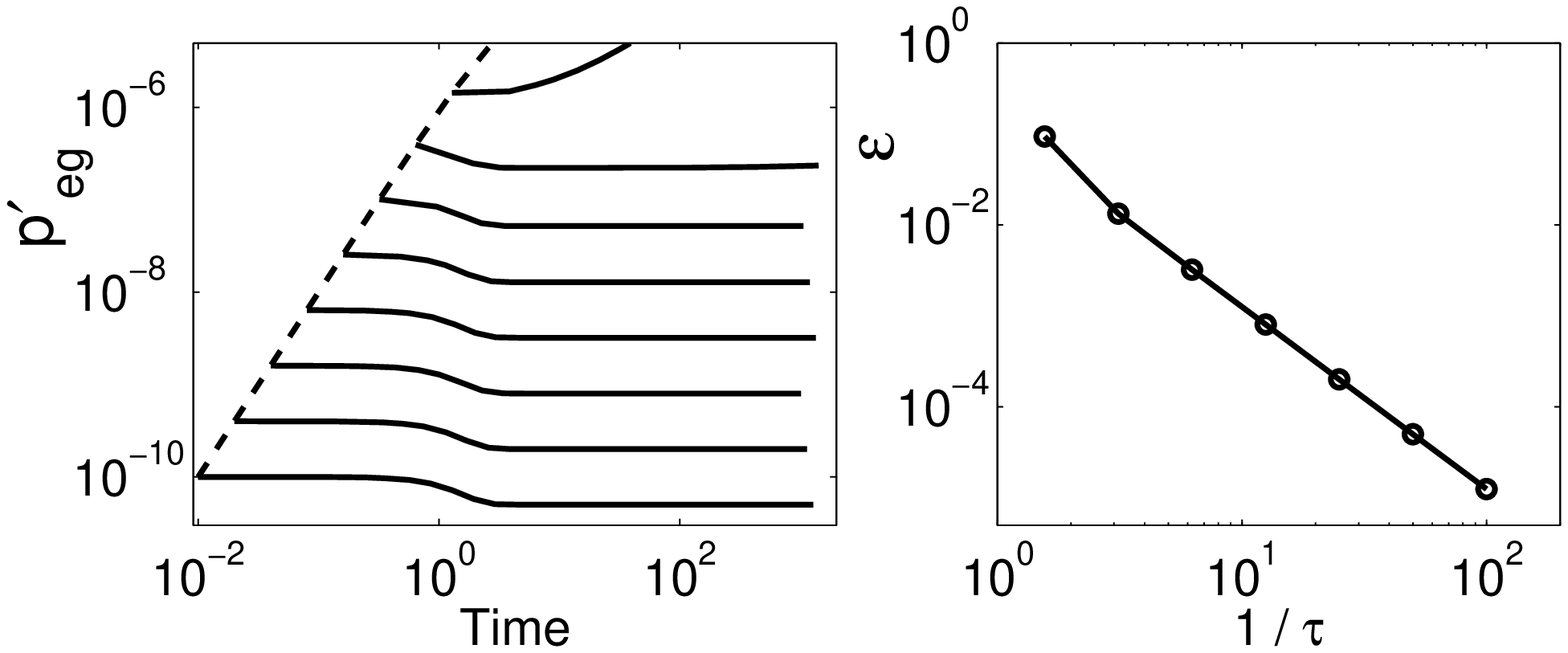}
\includegraphics[width=3.25in]{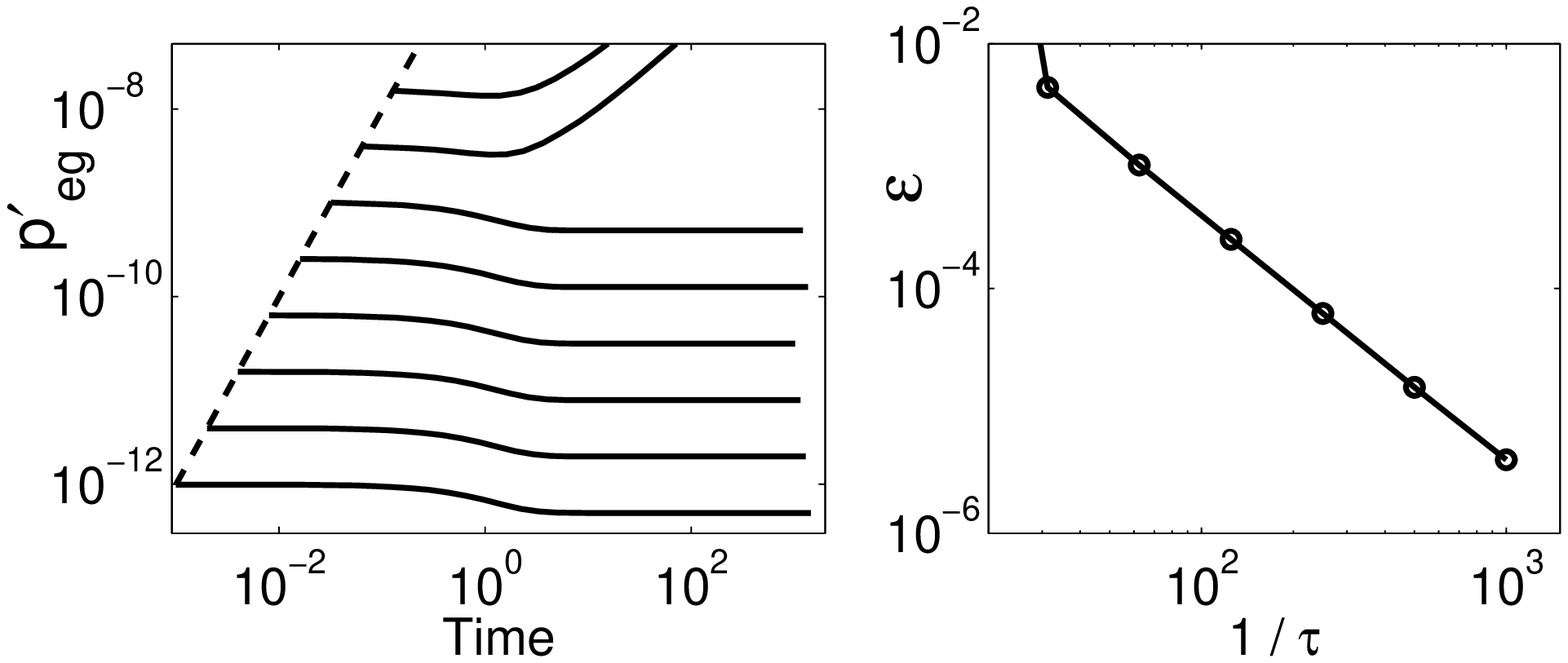}
\includegraphics[width=3.25in]{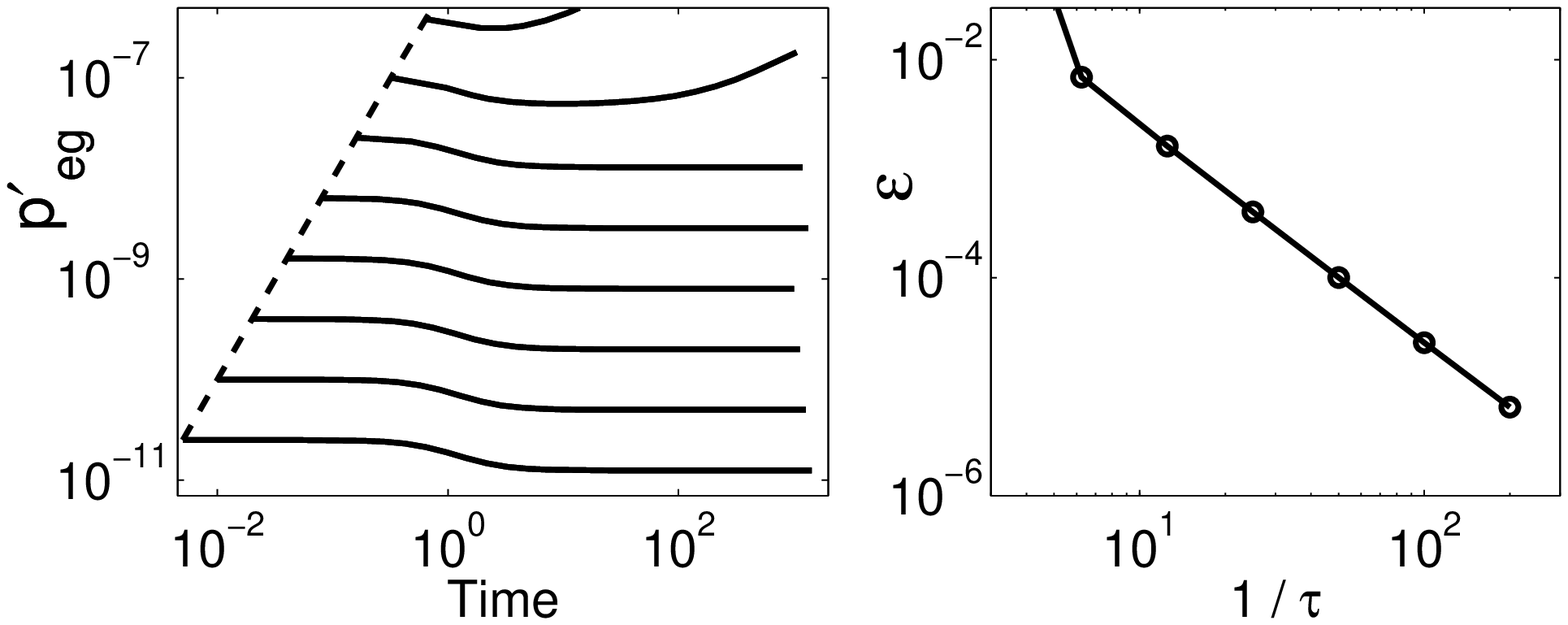}
\caption{Decay freezing (left column) and dependence of the differential
freezing value on pulse delay (right column) for the Gaussian (top row),
Lorentzian (middle row), and exponential (bottom row) spectrum. Only results of
odd number pulses are plotted for a better view. The dashed lines show decay
without control pulses. Crosses of dashed lines and solid lines in the left
column panels denote the specific pulse delay $\tau$.}
\label{fig:cse1biitau}   
\end{figure}

By inspecting Eq.~(\ref{eq:cse}), the controlled transition probability freezes
if $N\lambda \gg 1$ and the freezing value is
\begin{eqnarray}
\lim_{N\rightarrow \infty} p_{eg}' &=& {1 \over 2 }\int d\omega \rho(\omega)
{\nu^2 \over \Omega^2}
    \sin^2\Omega \tau {1 \over \sin^2\lambda},
\end{eqnarray}
where we have replaced the rapidly oscillating integrand $\sin^2(N\lambda)$
with its average $1/2$. The left column of Fig.~\ref{fig:cse1biitau} from
numerical calculation indeed shows that the transitions freeze at long times
for three different cases at small $\tau$. The smaller the $\tau$ is, the
smaller the freezing value is. Freezing of the transition probability is
equivalent to freezing of the survival probability of the excited state, $f =
1-p'_{eg}$.

Moreover, for small enough pulse delay $\tau$, we have $\sin^2\lambda \approx
1$ thus
\begin{eqnarray}
\lim_{N\rightarrow \infty} p_{eg}' &\approx & {1 \over 2 }\int d\omega
\rho(\omega) {\nu^2 \over \Omega^2}
    \sin^2\Omega \tau = {1\over 2} p_{eg}(\tau).
    \label{eq:qze}
\end{eqnarray}
The above relation indicates that the freezing value of the transition
probability at long times is one half of the free transition value at $t=\tau$
which is the pulse delay. Clearly, the road map to the quantum ZE is that the
decay rate becomes zero after initial pulses (decay freezing) and the freezing
value of the survival probability $f$ approaches 1 as $\tau$ decreases.

Define differential freezing value $\varepsilon (\tau) = p_{eg}'(N\tau=1000) -
{1\over 2} p_{eg}(\tau)$, which describes the difference of the transition
probability after many pulses and one half of the decay in the first pulse delay.
The right column of Fig.~\ref{fig:cse1biitau} shows that $\varepsilon$
decreases with increasing of $1/\tau$ (decreasing of $\tau$), confirming the
analytical results of Eq.~(\ref{eq:qze}).

In fact, decay freezing exists not only in the model system we consider in the
paper, but also exists in many other pulse-controlled systems, such as gated
semiconductor quantum dot~\cite{Zhang07a, Zhang07r, Zhang08, Lee08, Liu07},
spin-boson model~\cite{Viola98e, Faoro04}, and nuclear
spins~\cite{Haeberlen76}. The basic idea behind the decay freezing is that the
control pulses creates an effective preferred direction along which the decay
is frozen. In terms of Pauli matrix (c.f. Eq.~\ref{eq:HPauli}), the preferred
direction created by control pulses in the model we consider is $z$.

\section{Conclusion and discussion}
\label{sec:cd}

By investigating three models of coupling spectrum intensity (Gaussian,
Lorentzian and exponential), we demonstrate that a two-level system subjected
to many ideal $2\pi$ pulses exhibits both quantum Zeno and anti-Zeno effect,
depending on the relative position of $\omega_0$ to the peak position
$\omega_m$ of the spectrum and the pulse delay $\tau$. Instead of decreasing
the decay rate, the pulsed two-level system shows decay freezing after many
pulses at small $\tau$ and the freezing value of the survival probability of
the initial excited state approaches 1 (no decay) with decreasing $\tau$.

In this paper, all the spectrums have single peak and we observe only single
quantum Zeno and/or anti-Zeno region. Under some special circumstances where a
multiple peaks spectrum exists, one would expect multiple quantum Zeno and
anti-Zeno regions. We have also assumed that the spectrum of the structured
vacuum is time independent where the back action exerted on the vacuum by the
two-level system has been neglected. A full quantum version of the coupling
between the two-level system and the vacuum could possibly change the picture
of the controlled decay at long times but the short time behavior would be
intact in the weak coupling regime, because the back action is weak and needs a
long time to manifest its effect on the two-level system dynamics.

We consider only periodic pulse sequence ($\tau$ is fixed) in this
paper. In principle, other pulse sequences with varying $\tau$ may
also give similar quantum Zeno and anti-Zeno effect. They may even
have additional advantages~\cite{Uhrig07, Lee08}. In addition, the
strength and the duration of the $2\pi$ pulses are finite in
practice. One could minimize the finite pulse effect by employing
the phase alternation techniques~\cite{Slichter92} or the Eulerian
protocols~\cite{Viola04}.

\section{Acknowledgments}

W. Z. is grateful for many helpful discussions with V. V. Dobrovitski, S. Y.
Zhu, and T. Yu. This work was partially carried out at the Ames Laboratory,
which is operated for the U. S. Department of Energy by Iowa State University
under Contract No. W-7405-82 and was supported by the Director of the Office of
Science, Office of Basic Energy Research of the U. S. Department of Energy.
Part of the calculations was performed at the National High Performance
Computing Center of Fudan University.


\begin{thebibliography}{28}
\bibitem[{dip()}]{dipolenote}
\bibinfo{note}{We have assumed the dipolar interaction has a symmetry such that
  $\langle e| \vec d\cdot \vec E |g\rangle = \langle g| \vec d\cdot \vec E
  |e\rangle$, where $\vec d$ and $\vec E$ are the dipole operator of the atom
  and the electric field of the light, respectively.}
\bibitem{picturenote}
We adopt the Schr\"odinger picture instead of the usual interaction picture (e.g., Ref.~\cite{Agarwal01}) for our convenience. Both pictures are equivalent in principle.
\bibitem{mmnote}
\bibinfo{note}{For weak couplings, which are the case in the decay problem we
consider, high order nonlinear interaction between the two-level
atom and multiple electric field modes is neglected.}
\end{thebibliography}


\end{document}